\documentclass[aps,prd,showpacs,twocolumn,nofootinbib]{revtex4}
\usepackage{graphicx}
\usepackage{dcolumn}
\usepackage{bm}

\begin{document}

\def\be{\begin{equation}}
\def\ee{\end{equation}}
\def\bd{\begin{displaymath}}
\def\ed{\end{displaymath}}
\def\ba{\begin{eqnarray}}
\def\ea{\end{eqnarray}}
\def\lr{\leftrightarrow}
\def\nn{$n\bar n$ }
\def\qq{$q\bar q$ }
\def\cc{$c\bar c$ }
\def\ccbar{$c\bar c$ }
\def\x{\bf x}
\def\B{\rm B}
\def\D{\rm D}
\def\E{\rm E}
\def\F{\rm F}
\def\G{\rm G}
\def\H{\rm H}
\def\I{\rm I}
\def\J{\rm J}
\def\K{\rm K}
\def\L{\rm L}
\def\M{\rm M}
\def\P{\rm P}
\def\S{\rm S}
\def\T{\rm T}
\def\V{\rm V}
\title{\Large Higher Charmonia and X,Y,Z states with Screened Potential}
\author{
Bai-Qing Li$^{a,b}$ and Kuang-Ta Chao$^a$}

\affiliation{$^a$Department of Physics and State Key Laboratory of
Nuclear Physics and Technology, Peking University, Beijing 100871,
China; \\$^b$Department of Physics, Huzhou Teachers College, Huzhou
313000, China}

\date{\today}

\begin{abstract}
We incorporate the color-screening effect due to light quark pair
creation into the heavy quark-antiquark potential, and investigate
the effects of screened potential on the spectrum of higher
charmonium. We calculate the masses, electromagnetic decays, and E1
transitions of charmonium states in the  screened potential model,
and propose possible assignments for the newly discovered charmonium
or charmonium-like $"X,Y,Z"$ states. We find the masses of higher
charmonia with screened potential are considerably lower than those
with unscreened potential.  The $\chi_{c2}(2P)$ mass agrees well
with that of the Z(3930), and the mass of $\psi(4415)$ is compatible
with $\psi(5S)$ rather than $\psi(4S)$. In particular, the
discovered four $Y$ states in the ISR process, i.e.,
$Y(4008),~Y(4260),~Y(4320/4360),~Y(4660)$ may be assigned as the
$\psi(3S),~\psi(4S),~\psi(3D),~\psi(6S)$ states respectively. The
$X(3940)$ and $X(4160)$ found in the double charmonium production in
$e^+e^-$ annihilation may be assigned as the $\eta_c(3S)$ and
$\chi_{c0}(3P)$ states. Based on the calculated E1 transition widths
for $\chi_{c1}(2P)\to \gamma J/\psi$ and $\chi_{c1}(2P)\to \gamma
\psi(2S)$ and other results, we argue that the $X(3872)$ may be a
$\chi_{c1}(2P)$ dominated charmonium state with some admixture of
the $D^0\bar{D}^{*0}$ component. Possible problems encountered in
these assignments and comparisons with other interpretations for
these $X,Y,Z$ states are discussed in detail. We emphasize that more
theoretical and experimental investigations are urgently needed to
clarify these assignments and other interpretations.

\end{abstract}

\maketitle
\section{Introduction}
In recent years, a number of charmonium-like states, the so-called
$"X,Y,Z"$ mesons, have been found at B factories and other
experiments (for reviews see
e.g.~\cite{Swanson06:XYZ,Olsen-XYZ,zhu}). These states all lie above
the open charm (e.g. $D(^*)\overline{D(^*)}$) thresholds, and the
observed masses and decays make the identifications of these new
states very puzzling. Apart from the conventional charmonium, many
exotic candidates, such as molecules, tetra-quarks and
charmonium-hybrids, are suggested. However, despite many exciting
hints, none of the exotic states has been firmly established so far,
and more theoretical studies are needed to explain the existing data
and to confront new experimental tests. In the domain of
conventional charmonium, one of the main difficulties to assign the
new mesons as excited charmonia is that the observed masses do not
fit the predictions of potential models with linear confinement
potential (the quenched potential). However, the situation can be
more complicated than the simple potential model calculations by
including the coupled channel effects,  or the string breaking
effects. These effects could make the masses of higher charmonia
lower than the potential model predictions.  One distinct example is
the state $Z(3930)$ observed in the two-photon
process~\cite{Belle06:Z3930}, which is now identified with the $2P$
charmonium $\chi_{c2}'$, but its mass is about 40 MeV lower than the
prediction given by the quenched potential model~\cite{Barnes2005}.
Another example is the $X(3940)$ observed in double charmonium
production in $e^+e^-$ annihilation~\cite{Belle07:X3940}, which is
likely (though not necessarily) to be the $3S$ spin-singlet
charmonium state $\eta_{c}''$~\cite{Eichten06:X3940}, but this
assignment will imply that the mass of the $3S$ pseudoscalar
charmonium  is  smaller than the prediction given by the quenched
potential model by about 100 MeV~\cite{Barnes2005}. Therefore, it is
possible that the quenched potential model may overestimate the
masses of charmonia in the energy region well above the open charm
thresholds.

Although very successful in the prediction of the charmonium
spectrum below the open-charm threshold, it is well-known that the
quenched potential model, which incorporates a Coulomb term at short
distances and the linear confining interaction at large
distances~\cite{Eichten:1978tg,Godfrey:1985xj}, will not be reliable
in the domain beyond the open-charm threshold. This is because the
linear potential, which is expected to be dominant in this mass
region, will be screened or softened by the vacuum polarization
effects of dynamical fermions~\cite{Laermann:1986pu}. Unfortunately,
this screening effect has not been directly detected with the
standard Wilson loop technique in the unquenched lattice
calculations. The reason might be that the Wilson loop operator
almost decouples from the physical ground state at a large lattice
distance that consists of two disjoint strings~\cite{Bali01:PRept}.
In other words, to simulate the screening effect out with the Wilson
loop technique needs a very long lattice time, which might be too
far beyond the ability of the simulation at present. On the other
hand, the screening or the string breaking effect has been
demonstrated indirectly by the investigation of the mixing of a
static quark-antiquark string with a static light meson-antimeson
system in $n_f=2$ lattice QCD~\cite{Bali:2005fu}. This effect is
also confirmed by the calculations within some holographic QCD
models~\cite{Armoni08:Screening} recently. However, since the
simulations of lattice QCD still have large uncertainties and
difficulties in handling higher excited states, in order to
investigate the screening effects on the charmonium spectrum it
should be useful to improve the potential model itself to
incorporate the screening effect, and compare the model predictions
with experimental data. Such screened potential
models~\cite{chao1989,Chao:1992et,Ding:1993uy,Ding:1995he} were
proposed many years ago in the study of heavy quarkonium and heavy
flavor mesons, as well as light hadrons\cite{ZhangZY}. In recent
years the screened potential models have again been used to
investigate the heavy quarkonium spectrum and leptonic decay widths
~\cite{Segovia-etal:ScreenedPotential}. In addition to heavy
hadrons, the spectra of light hadrons have also been investigated
with the screened potential\cite{lighthadron}, and it is argued that
the large degeneracy observed in the excited meson spectrum by the
Crystal Barrel Collaboration in proton-antiproton annihilation in
the range 1.9-2.4 GeV\cite{Bugg04} may indicate the flattening of
the confinement potential due to the color screening
effects\cite{lighthadron}. On this point, it is important to further
examine experimentally whether or not the linear Regge trajectory of
the meson spectrum can hold for even higher excited mesons. More
complete data are needed in the future to clarify the issue of the
light meson spectrum regarding the color-screening effect. Although
no definite conclusion can be drawn at present, it is certainly
useful to study the color-screening effects on the mass spectra for
both heavy and light hadrons, especially for the newly discovered
higher excited states.

The effect of vacuum polarization due to dynamical quark pair
creation may also be compensated by that of the hadron loop induced
by virtual $D$ meson pairs in the so-called coupled-channel
model~\cite{Eichten:1978tg,Tornqvist84:CCM,Pennington07:CCM,Kalashnikova:2005ui},
for charmonium system. In Ref.~\cite{Li08:Couppled-vs-Screened}, we
compared the coupled-channel model with the screened potential model
in charmonium spectrum in the mass region below 4 GeV. With the same
quenched limit, the two models are found to have similar global
features. It is not surprising since, in the quark-meson duality
picture, the two models may describe roughly the same effects. In
practice, calculations with the screened potential model are simpler
to deal with, though detailed predictions for the spectrum can be
somewhat different from the coupled-channel model
~\cite{Li08:Couppled-vs-Screened}.

In this paper, we calculate the mass spectrum of the charmonium
especially the higher charmonium using a non-relativistic
Schr\"{o}dinger equation with the Coulomb potential plus a screened
linear potential which is nearly the same as that
in~\cite{chao1989,Ding:1993uy} but with slightly different
parameters~(see Sec.~II for details). Spin-dependent interactions
are considered perturbatively. With one additional screening
parameter $\mu$, the model predicts that the masses of higher
charmonium are lowered, compared with the quenched linear potential,
and this mass suppression tends to be strengthened when the
charmonium states vary from lower levels to higher ones.
For instance, we find that the calculated mass of $\chi_{c2}^{'}$ is
3937 Mev, fitting well the experimental value of
$Z(3930)$~\cite{Belle06:Z3930}, and the mass of $\psi(5S)$ is close
to the observed $\psi(4415)$. Consequently, this mass spectrum will
leave more room for new assignments for some of the observed $X,Y,Z$
states in the charmonium family. These possible assignments will be
suggested and discussed in detail in this paper.

In the following, we first introduce the screened potential model in
Sec.II, and then study some decay and transition processes of
charmonia in Sec.~III.  In Sec.~IV we will discuss possible
assignments for the observed charmonium(-like) states. A summary
will be given in Sec.~V.

%

\section{The screened potential model}
\label{Themodel} As a minimal model describing the charmonium
spectrum we use a non-relativistic potential model with screening
effect being considered~\cite{chao1989,Ding:1993uy}. We use a
potential as
\begin{equation}
 V_{scr}(r) = V_{V}(r) + V_{S}(r) ,
\end{equation}
\begin{equation}
 V_{V}(r) = -\frac{4}{3}\frac{\alpha_C}{r},
\end{equation}
\begin{equation}
 V_{S}(r) = \lambda (\frac{1-e^{-\mu r}}{\mu}),
\end{equation}
where $\mu$ is the screening factor which makes the long range
scalar part of $V_{scr}(r)$ flat when $r \gg \frac{1}{\mu}$ and
still linearly rising when $r \ll \frac{1}{\mu}$, $\lambda$ is the
linear potential slope, and $\alpha_C$ is the coefficient of the
Coulomb potential. Figure 1. shows the screened potential departure
from linear in large length with the parameters given in
(\ref{para}).


\begin{figure}[t]
\begin{center}
\vspace{0cm} \hspace*{0cm}
\scalebox{0.5}{\includegraphics[width=16cm,height=9cm]{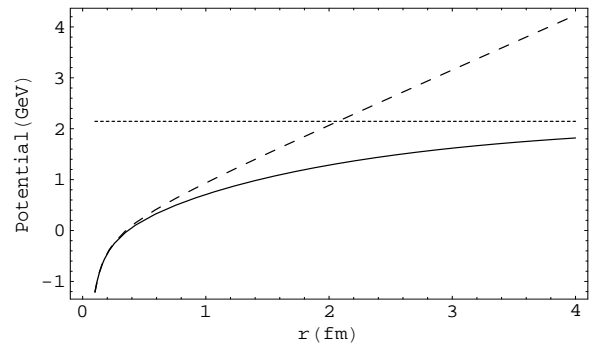}}
\end{center}
\vspace{0cm}\caption{Comparison of the unscreened potential $V(r)$
(dashed line) and the screened potential $V_{scr}$ (solid line) for
$r=0.1\mbox{-}4$ fm with parameters taken from (\ref{para}). The
asymptotic limit of $V_{scr}$ is shown by the dotted
line.}\label{Fig:Potential}
\end{figure}

For hyperfine interactions, we only consider the spin-dependent
interactions which include three parts as follows.

The spin-spin contact hyperfine interaction is
\begin{equation}
\label{Hss}
 H_{SS} =\frac{32\pi\alpha_C}{9 m_c^2}\, \tilde \delta_{\sigma}(r)\, \vec
{S}_c \cdot \vec {S}_{\bar c}\, ,
\end{equation}
where $\tilde \delta_{\sigma}(r)$ is usually taken to be $\delta
(\vec r\,)$ in nonrelativistic potential models, but here we take
$\tilde \delta_{\sigma}(r) = (\sigma/\sqrt{\pi})^3\, e^{-\sigma^2
r^2}$ as in Ref.\cite{Barnes2005} since it is an artifact of an
O$(v_c^2/c^2)$ expansion of the T-matrix~\cite{Barnes1982} in a
range comparable to $1/m_c$.

The spin-orbit term and the tensor term take the common forms:
\begin{equation}
\label{Hls}
 H_{LS} = \frac{1}{2m_{c}^{2}r} (3V_{V}^{'}(r)- V_{S}^{'}(r)) \vec
 {L} \cdot \vec {S},
\end{equation}
 and
\begin{equation}
\label{Ht}
 H_{T} = \frac{1}{12m_{c}^{2}}(\frac{1}{r}V_{V}^{'}(r)-V_{V}^{''}(r)) T.
\end{equation}
 These spin-dependent interactions are dealt with perturbatively. They are
diagonal in a $|J,L,S>$ basis with the matrix elements
\begin{equation}
 < \vec {S}_c \cdot \vec {S}_{\bar c} > =
\frac{1}{2} S^{2}-\frac{3}{4},
\end{equation}
\begin{equation}
< \vec {L} \cdot \vec {S} > = [J(J+1) - L(L+1) - S(S+1)]/2
\end{equation}
and the tensor operator T has nonvanishing diagonal matrix elements
only between $L>0$ spin-triplet states, which are
\begin{eqnarray}
<^3L_J |T| ^3L_J> =\left\{\begin{array}{ll}
   -\frac{L}{6(2L + 3)}\,,&J = L + 1\\\rule{0cm}{0.7cm}\frac{1}{6}\,,&J = L\\\rule{0cm}{0.7cm}-\frac{(L + 1)}{6(2L - 1)}\,,&J = L-1.
\end{array}\right.
\end{eqnarray}
 For the model parameters, we take
\begin{eqnarray}
 \label{para}
 \alpha_C=0.5007,~~~~~~\lambda=0.21GeV^2, \nonumber\\
\mu=0.0979GeV,~~~~~~~\sigma=1.362GeV,\nonumber\\
m_c=1.4045GeV,~~~~~~~\alpha_{S}=0.26,
\end{eqnarray}
where $\alpha_C\approx \alpha_{s}(m_c v_c)$ and $\alpha_{S}\approx
\alpha_{s}(2m_c)$ are essentially the strong coupling constants but
at different scales. The former is for large distances and used to
determine the spectrum while the latter is for short-distances and
used for QCD radiative corrections in charmonium decays (see below
in next section). Here $\mu$ is the characteristic  scale for color
screening, and $1/{\mu}$ is about 2 $fm$, implying that at distances
larger than $1/{\mu}$ the static color source in the $c\bar c$
system gradually becomes neutralized by the produced light quark
pair, and string breaking emerges. With these values of the
parameters for the potential, we can calculate the spectrum of the
charmonium system. The results are shown in
Table~\ref{spectrum_table}. For comparison, we also list the
experimental values~\cite{PDG08} and those predicted by the quenched
potential model~\cite{Barnes2005} in Table~\ref{spectrum_table}.

\section{Some decay processes}

\subsection{Leptonic decays} The electronic decay width of the vector meson
is given by the Van Royen-Weisskopf formula~\cite{VanRoyen:1967nq}
with QCD radiative corrections taken into
account~\cite{Barbieri:1979be}.
\begin{equation}
\label{Swave:ee}
\Gamma_{ee}(nS)=\frac{4\alpha^{2}e_{c}^{2}}{M_{nS}^{2}}
|R_{nS}(0)|^{2} (1-\frac{16}{3}\frac{\alpha_{S}}{\pi}),
\end{equation}
\begin{equation}
\label{Dwave:ee} \Gamma_{ee}(nD) =
\frac{25\alpha^{2}e_{c}^{2}}{2M_{nD}^{2}m_{Q}^{4}}
|R_{nD}^{''}(0)|^{2} (1-\frac{16}{3}\frac{\alpha_{S}}{\pi}),
\end{equation}
where $M_{nS}(M_{nD})$ is the mass for $nS(nD)$, $e_{c}=\frac{2}{3}$
is the c quark charge in unit of electron charge, $\alpha$ is the
fine structure constant, $R_{nS}(0)$ is the radial S wave-function
 at the origin, and $R_{nD}^{''}(0)$ is the second derivative of the
radial D wave-function at the origin.

Combined with the parameters~(\ref{para}), we get the results listed
in Table~\ref{tab:ee}.

\subsection{Two-photon decays} In the nonrelativistic limit, the
two-photon decay widths of ${}^1S_{0}$, ${}^3P_{0}$, and ${}^3P_{2}$
can be written as~\cite{kmrr}
\begin{eqnarray}
\label{2gs0}
 \Gamma^{NR}({}^1S_0\to
 \gamma\gamma)&=&\frac{3\alpha^2e_c^4|R_{nS}(0)|^2}{m_c^2}\,,\\
\label{2gp0}
\Gamma^{NR}({}^3P_0\to\gamma\gamma)&=&\frac{27\alpha^2e_c^4|R'_{nP}(0)|^2}{m_c^4}\,,\\
\label{2gp2}
\Gamma^{NR}({}^3P_2\to\gamma\gamma)&=&\frac{36\alpha^2e_c^4|R'_{nP}(0)|^2}{5m_c^4}.
\end{eqnarray}

The first-order QCD radiative corrections to the two-photon decay
rates can be accounted for as~\cite{kmrr}
\begin{eqnarray}
\Gamma({}^1\!S_0\to\gamma\gamma)&=&\Gamma^{NR}({}^1\!S_0\to
\gamma\gamma)[1+\frac{\alpha_S}{\pi}(\frac{\pi^2}3-\frac{20}3)]\,,\rule{0.5cm}{0cm}\\
\Gamma({}^3\!P_0\to \gamma\gamma)&=& \Gamma^{\rm NR}({}^3\!P_0\to
\gamma\gamma)[1+\frac{\alpha_S}{\pi}
(\frac{\pi^2}3-\frac{28}9)]\,,\rule{0.5cm}{0cm}\\
\Gamma({}^3\!P_2\to \gamma\gamma)&=& \Gamma^{\rm NR}({}^3\!P_2\to
\gamma\gamma)[1 -\frac{16}3\frac{\alpha_S}{\pi} ]\,.
\end{eqnarray}

We can see that $\Gamma({}^1\!S_0\to\gamma\gamma)\propto
|R_{nS}(0)|^2$, which are sensitive to the details of potential near
the origin. So we take
\begin{equation}
\Gamma({}^1\!S_0\to\gamma\gamma)\longrightarrow
\frac{\Gamma({}^1\!S_0\to\gamma\gamma)}{\Gamma_{ee}(nS)}\Gamma^{expt}_{ee}(nS)
\end{equation}
to eliminate this uncertainty.

In the nonrelativistic limit, we can also replace $m_{c}$ by M/2,
where M is the mass of the corresponding  charmonium state. The
results are listed in Table~\ref {twophoton}.

\subsection{E1 transitions}

For the E1 transitions within the charmonium system, we use the
formula of Ref.~\cite{Kwong:1988ae}:
\begin{eqnarray}
&&\Gamma_{\rm E1}( {\rm n}\, {}^{2{\S}+1}{\rm L}_{\J} \to {\rm n}'\,
{}^{2{\S}'+1}{\rm L}'_{{\J}'} + \gamma)\nonumber\\
&&=\frac{4}{3}\, C_{fi}\, \delta_{{\S}{\S}'} \, e_c^2 \, \alpha \,
|\,\langle f | \, r \, |\, i \rangle\, |^2 \, {\rm E}_{\gamma}^3 \,
\end{eqnarray}
 where E$_{\gamma}$ is the emitted photon energy.

The spatial matrix element
\begin{equation}
<f|r|i>=\int_{0}^{\infty}R_{f}(r)R_{i}(r)r^{3}dr\,,
\end{equation}
involves the initial and final state radial wave functions, and the
angular matrix element $C_{fi}$ is
\begin{equation}
C_{fi}=\hbox{max}({\L},\; {\L}')\; (2{\J}' + 1) \left\{ { {{\L}'
\atop {\J}} {{\J}' \atop {\L}} {{\S} \atop 1}  } \right\}^2 .
\end{equation}

Our results are listed in Table~\ref{E1rad}. The widths calculated
by the zeroth-order wave functions are marked by $SNR_0$ and those
by the first-order relativistically corrected wave functions are
marked by $SNR_1$.

For the first-order relativistic corrections to the wave functions,
we include the spin-dependent part of
(\ref{Hss}),(\ref{Hls}),(\ref{Ht}) and the spin-independent part
as~\cite{Miller:1982ca}
\begin{eqnarray}
H_{SI}&=&-\frac{\vec{P}^4}{4m_c^3}+\frac{1}{4m_c^2}\bigtriangledown ^2V_V(r)\nonumber\\
&&-\frac{1}{2m_c^2}\left\{\left\{
\vec{P}_1\cdot V_V(r) \Im \cdot \vec{P}_2 \right\}\right\}\nonumber\\
    &&+\frac{1}{2m_c^2}\left\{\left\{\vec{P}_1\cdot \vec{r}\frac{V_V^{'}(r)}{r}\vec{r}\cdot \vec{P}_2
    \right\}\right\},
\end{eqnarray}
where $\vec{P}_1$ and $\vec{P}_2$ are momenta of c and $\bar c$
quarks in the rest frame of charmonium, respectively, which satisfy
$\vec{P}_1=-\vec{P}_2=\vec{P}$, $\Im$ is the unit second-order
tensor, and $\{\{\quad\}\}$ is the Gromes's notation
\begin{equation}
\{\{\vec{A}\cdot \Re\cdot
\vec{B}\}\}=\frac{1}{4}(\vec{A}\vec{B}:\Re+\vec{A}\cdot \Re
\vec{B}+\vec{B}\cdot \Re \vec{A}+\Re:\vec{A}\vec{B}),
\end{equation}
where $\Re$ is any second-order tensor.

Note that we do not include the contributions from the scalar
potential in $H_{SI}$ since it is still unclear how to deal with the
spin-independent corrections arising from the scalar potential
theoretically.

The results of non-relativistic linear potential
model~\cite{Barnes2005} marked by NR are listed for comparison. Our
results for the E1 transition widths of $SNR_0$ are slightly larger
than those of the NR's mainly due to the relativistic phase space
factor of $\E_f^{(c\bar c)}/\M_i^{(c\bar c)}$ involved in the NR's
in~\cite{Barnes2005} , where $\E_f^{(c\bar c)}$ is the final state
charmonium total energy, and $\M_i^{(c\bar c)}$ is the initial state
charmonium mass. Both $SNR_0$ and NR's results of E1 transitions are
larger than experimental values, but we see that in $SNR_1$  the
predicted widths get decreased and fit the experimental values
rather well as long as the first-order relativistic corrections to
the wave functions are included.

\section{Discussions}
\label{Discussion}

\subsection{$Z(3930)$}
The $Z(3930)$ was discovered by the Belle
Collaboration~\cite{Belle06:Z3930} as an enhancement in the
$D\bar{D}$ invariant mass near 3.93~GeV/$c^2$ in the $\gamma \gamma$
collision with the statistical significance of $5.3\sigma$. Results
for the mass, width, and product of the two-photon decay width times
the branching fraction to $D\bar{D}$ are:
\begin{eqnarray}
M(Z(3930))&=&3929 \pm 5 \pm 2~\mbox{MeV},\label{m:Z3930}\\
\Gamma(Z(3930))& = &29 \pm 10\pm 2~\mbox{MeV},\\
\hspace{-5mm}\Gamma_{\gamma\gamma}{\cal B}(Z(3930)\to D\bar{D})&=&
0.18 \pm 0.05 \pm 0.03~\mbox{KeV},\label{product-width:Z3930}
\end{eqnarray}
 respectively. The production rate and the angular distribution in
the $\gamma\gamma$ center-of-mass frame suggest that this state is
the previously unobserved $\chi_{c2}(2P)$~\cite{Belle06:Z3930}.

The mass of $\chi_{c2}(2P)$ in our model is 3937 MeV, which is
consistent with the experimental value in (\ref{m:Z3930}), whereas
the mass in the quenched potential model~\cite{Barnes2005} is larger
than the experimental one by 40-50 MeV (see
Tabel~\ref{spectrum_table}). As we have mentioned, this is one of
our motivations to reexamine the charmonium spectrum in the screened
potential model.

The open-charmed decays of $\chi_{c2}(2P)$ were studied in
Ref.~\cite{Swanson06:XYZ} (the mass was set to be 3931 MeV), and the
total width is predicted to be 35 MeV with the branching ratio
${\cal B}(Z(3930)\to D\bar{D})\simeq74\%$. Together with the data in
(\ref{product-width:Z3930}), one can get the two-photon decay width
of $Z(3930)$ as about 0.16-0.33 KeV, which is consistent with the
predicted value 0.23~KeV for $\chi_{c2}(2P)$ in our mode (see
Tabel~\ref{twophoton}).

%
%

\subsection{$\psi(3770), \psi(4040), \psi(4160), \psi(4415)$}

Before the discovery of $X(3872)$, there are only four
well-established charmonium states above the $D\bar{D}$ threshold.
They are $\psi(3770)$, $\psi(4040)$, $\psi(4160)$ and $\psi(4415)$.
They all have quantum number $J^{PC}=1^{--}$. Conventionally, they
were assigned as mainly $\psi(1^{3}D_1)$, $\psi(3^{3}S_1)$,
$\psi(2^{3}D_1)$ and $\psi(4^{3}S_1)$, respectively. These
assignments are consistent with the predictions given by the
quenched potential models. However, in our screened potential model,
the $\psi(4415)$ should be assigned as $\psi(5^{3}S_1)$ (see
Table~\ref{spectrum_table}).

The experimental di-electron width of $\psi(3770)$ is
$0.259\pm0.016$ KeV, larger than that expected for a pure D-wave
state, which is probably due to mixing with the $2^3S_1$ state
induced  by the coupled channel effects as well as by the tensor
force. In the following we assume $\psi^{'}\equiv\psi(3686)$ and
$\psi^{''}\equiv\psi(3770)$ to be admixture of
$2^{3}S_{1}-1^{3}D_{1}$ charmonium states and
\begin{eqnarray}
&|\psi^{'}\rangle&=~~\, |2{}^3{\rm S}_1\rangle \cos\theta+
 |1{}^3{\rm D}_1\rangle\sin\theta,\\
&|\psi^{''}\rangle&=-
 |2{}^3{\rm S}_1\rangle\sin\theta+ |1{}^3{\rm
D}_1\rangle\cos\theta ,
\end{eqnarray}
where $\theta$ is the mixing angle. This angle can be estimated by
comparing the experimental values of the di-electron widths of
$\psi(3770)$ and $\psi(3686)$ with theoretical predictions for pure
$1^3D_1$ and $2^3S_1$ states (see Table~\ref{tab:ee}). The result is
given by
\begin{equation}
\theta\approx -12^\circ\quad \mbox{or}\quad \theta\approx 25^\circ.
\end{equation}
Thereinto, $\theta=25^\circ$ is not compatible with the width of
$\psi^{'}\rightarrow\chi_{c0}\gamma$, so we take $\theta=-12^\circ$.
Note that $\theta\approx-12^\circ$ is consistent with the results of
Refs.~\cite{Kuang:1989ub,Ding:1991vu,Eichten:1979ms,Tornqvist84:CCM}.

The S-D mixing may be more serious for $\psi(4160)$
 (with $\Gamma_{ee}(\psi(4160))=0.83\pm$ 0.07~KeV\cite{PDG08}), if one assign it as the
$2^3D_1$ state, since its observed di-electron width is comparable
to that of $\psi(4040)$
 (with $\Gamma_{ee}(\psi(4040))=0.86\pm$ 0.07~KeV\cite{PDG08}). Extracted from the ratio of
di-electron widths of $\psi(4160)$ and $\psi(4040)$, the mixing
angle between $2^3D_1$ and $3^3S_1$ can be as large as $-37^\circ$
in our model and $-35^\circ$ in
Refs.~\cite{Chao08:X4160,Badalian08:S-D-mixing}. Moreover, there
could also be mixing between $2D-4S$ states. This indicates that the
observed di-electron widths for higher charmonia (above the open
charm threshold) can be altered by the S-D mixing effects due to
coupling to the decay channels. Moreover, their masses can also be
modified by the S-D mixing.

The open-charmed decays of these states are not studied in this
paper. However, one may expect that the corresponding calculations
and results in our model will be similar to those in the quenched
potential model. In Ref.~\cite{Barnes2005}, the authors evaluated
the open-charmed decay widths of these states using the $^3P_0$
quark pair creation model, and the results are consistent with
experimental measurements both for the total widths and the decay
patterns. Especially, they predicted that the main decay modes of
$\psi(4415)$ (as $\psi(4S)$ in their model) are $D\bar{D}_1$ and
$D\bar{D}_2^*$~\footnote{In this paper, inclusion of the charge
conjugate mode is always implied.}, and the $D\bar{D}_2^*$ mode was
confirmed by Belle's measurement~\cite{Belle08:psi4415-DD2}
recently. The lesson from both the theoretical calculation and the
experimental measurement is that the higher excited charmonia tend
to decay into excited charm mesons rather than the S-wave charm
meson pairs. This might be due to relativistic suppression and the
node structures of the wave functions of the higher excited states.

\subsection{$Y(4008), Y(4260), Y(4320/4360), Y(4660)$ and $X(4630)$}

The vector state $Y(4260)$ was first discovered by the BaBar
Collaboration~\cite{BaBar05:Y4260} as a relative narrow peak around
4260 MeV in the $J/\psi~\pi^{+}\pi^{-}$ distribution in the initial
state radiation (ISR) process $e^+e^- \to \gamma_{\rm ISR}
J/\psi~\pi^{+}\pi^{-}$. This state was also observed by
CLEO~\cite{CLEO06:Y4260} and Belle~\cite{Belle07:Y4008&4260}, and
there is also a relative broad structure near 4.05 GeV, the
so-called $Y(4008)$, in the Belle data. Recently,
BaBar~\cite{BaBar08:Y4260-updata} updated their measurement and gave
the mass and width of $Y(4260)$
which are consistent with Belle's measurements
~\cite{Belle07:Y4008&4260}. However, Babar has not yet seen the
structure around 4.05 GeV. The mass and width of the broad structure
Y(4008) are given by\cite{Belle07:Y4008&4260}
\ba M(Y(4008))&=&4008\pm 40^{+114}_{-28}~\mbox{MeV},\label{m:Y4008}\\
\Gamma(Y(4008))&=&226\pm 44\pm 87~\mbox{MeV}.\label{Gamma:Y4008} \ea
The averaged mass and width of Y(4260) are given by\cite{PDG08}
\ba M(Y(4260))&=&4263^{+8}_{-9}~\mbox{MeV},\label{m:Y4260}\\
\Gamma(Y(4260))&=&95\pm 14~\mbox{MeV}.\label{Gamma:Y4260} \ea
BaBar also found a structure around 4.32 GeV~\cite{BaBar07:Y4325} in
the ISR process $e^+e^- \to \gamma_{\rm ISR}  \psi' \pi^{+}\pi^{-}$
with \ba
M(Y(4320))&=&4324\pm24~\mbox{MeV},\\
\Gamma(Y(4320))&=&172\pm33~\mbox{MeV},
 \ea
while in the same process the Belle Collaboration observed two
relative narrow peaks around 4.35 GeV with
 \ba
M(Y(4360))&=&4361\pm9\pm9~\mbox{MeV},\\
\Gamma(Y(4360))&=&74\pm15\pm10~\mbox{MeV},
 \ea
 and 4.66 GeV~\cite{Belle07:Y4350&4660} with
 \ba
M(Y(4660))&=&4664\pm11\pm5~\mbox{MeV},\\
\Gamma(Y(4660))&=&48\pm15\pm3~\mbox{MeV}.
 \ea

Aside from the broad structure $Y(4008)$, which might be related to
the $\psi(3S)$ state $\psi(4040)$ and even the $\psi(2D)$ state
$\psi(4160)$, the other three $Y$-states are considered to be
difficult to assign as conventional charmonia since their masses are
inconsistent with those predicted by the quenched potential
models~\cite{Barnes2005,Ding08:Y4350&4660}, and there are even no
enough unassigned states in the charmonium spectrum to accommodate
them. As a consequence, these $Y$-states are interpreted totally or
partly as exotic states, such as $c\bar cg$
hybrid~\cite{Zhu:Y4260:hybrid}, $cq\bar c\bar q$ tetra-quark
state~\cite{Maiani:Y4260:4quark,Ebert08,Nielsen09},
bayonium~\cite{Qiao:Ystates:bayonium} and
molecule~\cite{Guo08:Y4660:molecule}. As the only exception, in
Ref.\cite{F.J. Llanes-Estrada} the Y(4260) was interpreted as the
$\psi(4S)$ charmonium. Recently, in Ref. \cite{Ding08:Y4350&4660}
the authors assigned $\psi(3D)$ and $\psi(5S)$ to the $Y(4325/4360)$
and $Y(4660)$ states, although the predicted masses are higher than
the experimental values by 50 to 100 MeV.

The situation changes greatly in our screened potential model. In
the earlier calculations\cite{Ding:1993uy,Ding:1995he}, the mass of
$\psi(4S)$ was just around 4260 MeV, the mass of Y(4260). Since the
predicted higher chamonium spectrum is compressed in the screened
potential model, there is enough space to accommodate these
$Y$-states. Specifically, in the model of this paper the masses of
$\psi(4S)$, $\psi(3D)$ and $\psi(6S)$ (see
Table~\ref{spectrum_table}) are predicted to be 4273, 4317 and 4608
MeV, which are roughly compatible with the observed masses of
$Y(4260)$, $Y(4325/4360)$ and $Y(4660)$, respectively. The small
mass discrepancies between theoretical predictions and experimental
data may be either due to the experimental errors or, more likely,
due to the theoretical uncertainties, especially the complicated S-D
mixing effects, such as the mixing among $4S$, $3D$ and $5S$ states.

The di-electron widths of the pure $4^3S_1$, $3^3D_1$ and $6^3S_1$
states in our model are 0.97, 0.044 and 0.49 and KeV, respectively.
Experience about the large S-D mixing (especially between
$\psi(4040)$ and $\psi(4160)$) discussed in the last subsection
tells us that the large S-D mixing may change these di-electron
widths to moderate values. Assuming the di-electron widths of
$Y(4260)$, $Y(4325/4360)$ and $Y(4660)$ to be all about 0.4 to 0.5
KeV, one can extract the branching ratios
$\mathcal{B}(Y\to\psi(\psi')~\pi^+\pi^-)$ from the experimental
measurements~\cite{Belle07:Y4008&4260,Belle07:Y4350&4660}. They are
all about $1\mbox{-}2\%$ and the corresponding widths
$\Gamma(Y\to\psi(\psi')~\pi^+\pi^-)=1\mbox{-}2$ MeV. These widths
look too large compared with $\Gamma(\psi'\to
J/\psi~\pi^+\pi^-)\sim100$ KeV~\cite{PDG08}. However, once the
higher charmonium lies well above the open charm threshold, the
di-pion transition rate may be enhanced dramatically by final state
interactions between charmed mesons, which are produced in the decay
of the charmonium. This is similar to the case in the $\Upsilon(5S)$
di-pion transitions~\cite{Meng:Y5S:dipion-transition}, where the
rescattering effects between $B^{(*)}-\bar B^{(*)}$ mesons are
expected to play a crucial role in enhancing the di-pion transition
rates of $\Upsilon(5S)$ into $\Upsilon(1S,2S)$. Furthermore, if the
broad structure $Y(4008)$ is indeed due to $\psi(4040)$ as well as
$\psi(4160)$, we will have a clear experimental hint for the large
di-pion transition rates of 3S as well as 2D charmonium, which are
well above the $D\bar D$ threshold.

It is useful to emphasize that for all the four newly discovered $Y$
states the
measurements\cite{BaBar05:Y4260,CLEO06:Y4260,Belle07:Y4008&4260,BaBar08:Y4260-updata,Belle07:Y4008&4260,Belle07:Y4350&4660}
\ba \Gamma_{ee}(Y)\times B(Y\to \psi \pi^+\pi^-)\approx
O(10~eV),\label{YeeBR} \ea where $\psi$ means $J/\psi$ or
$\psi(2S)$, and the measured total widths of about $O(100~MeV)$
altogether may imply that they have similar properties. Therefore a
coherent interpretation for these four $Y$ states is needed. A
likely explanation is that they are conventional charmonium states,
though other interpretations are also possible and even more
interesting. Note that in the hybrid scenario if one of $Y$ states,
e.g., the Y(4260) is the $1^{--}$ hybrid, then one needs to
understand why the others, which can no longer be accommodated as
$1^{--}$ hybrids in this mass region, have similar properties to the
Y(4260). On the other hand, there are some considered difficulties
to assign $Y$ states as conventional charmonia, and the most serious
one to assign Y(4260) as the 4S-dominated charmonium seems to be the
observed dip rather than a peak in the $R$ value scanned in $e^+e^-$
annihilation~\cite{PDG08} around Y(4260) (this difficulty is common
to all resonance interpretations of Y(4260)). A possible explanation
for the dip is the destructive interference between the continuum
and the resonance. If without any resonance in this region, the
continuum contribution should be generally smooth.
Another difficulty is the nonobservation (not a peak but a dip
observed) of the decay modes $D\bar D,~D\bar {D^*},~D^*\bar {D^*}$
at the Y(4260) (for a recent report by BaBar see
Ref.\cite{BaBar09:4260}). The dip in the observed charmed meson
pairs is probably related to the dip in the $R$ value, since the
latter is the measurement of the hadron production cross section in
$e^+e^-$ annihilation.  The dip in $R$ is just the reflection of the
dip in the resonance decays to hadrons (only the charmed hadrons are
relevant here). They may all be caused by the interference effects.
Moreover, the above mentioned difficulties are not only for the
charmonium assignment but also for other interpretations of Y(4260).
One needs to understand the dip in $R$ around Y(4260) if one tries
to interpret Y(4260) as a resonance no matter which kind of
resonance it is. Nevertheless, it is instructive to search for decay
modes involving the P-wave charm mesons e.g. $D_1\bar D$ and other
higher charm mesons, apart from the S-wave charm meson pairs,
because the higher charmonium may prefer decays to higher charm
mesons or multi-mesons, due to the form factor suppression with
higher momentum released in decays to lower charm mesons, and also
due to the node structure of higher charmonium state. In this
regard, we note that a main decay mode of the $\psi(4415)$ is
$D_2\bar D$~\cite{Belle08:psi4415-DD2}.


Very recently, Belle reported a new vector state
$X(4630)$~\cite{Belle08:X4630} which was found as a threshold
enhancement in the $\Lambda_c^+\Lambda_c^-$ distribution in the ISR
process $e^+e^- \to \gamma_{\rm ISR} \Lambda_c^+\Lambda_c^-$. The
mass and width are fitted to be
\ba M(X(4630))&=&4634_{-7-8}^{+8+5}~\mbox{MeV},\label{m:X4630}\\
\Gamma(X(4630))&=&92_{-24-21}^{+40+10}~\mbox{MeV},\label{Gamma:X4630}
\ea
which are roughly in agreement with those of $Y(4660)$. Assuming
that $X(4630)$ is the same state as the $Y(4660)$ ($\psi(6S)$ in our
model) with the di-electron width of about 0.5 KeV, one can extract
the partial width
\be
\Gamma(X(4630)\to\Lambda_c^+\Lambda_c^-)\sim10~\mbox{MeV},\label{Gamma:X4630}
\ee
i.e., the branching ratio of about $10\%$. Such a large baryonic
decay width certainly deserves further studying.

To sum up for the discussion in this subsection, our assignments for
these newly discovered $Y$ states appear to be consistent with the
4S, 3D, 6S charmonium mass spectrum predicted by the screened
potential model,   and other properties may also be understood in
these charmonium interpretations. But the issue is far from being
conclusive, and many theoretical and experimental investigations are
apparently needed to clarify these assignments with other more
interesting interpretations such as hybrids and tetraquarks.

\subsection{$X(3872)$}

The $X(3872)$ was first observed by Belle~\cite{Belle0309032} in the
$J/\psi~\pi^{+}\pi^{-}$ invariant mass distribution in
$B^{+}\rightarrow K^{+} J/\psi~\pi^{+}\pi^{-}$ decay as a very
narrow peak ($\Gamma_X<2.3$ MeV) around 3872 MeV. The mass of
$X(3872)$ in the $J/\psi~\pi^+\pi^-$ mode was recently updated by
CDF Collaboration~\cite{CDF08:X3872:mass} as
\be M(X(3872))=3871.61\pm0.16\pm0.19~\mbox{MeV},\label{M:X3872} \ee
which is very close to the $D^0\bar{D}^{*0}$ threshold
$m(D^0\bar{D}^{*0})=3871.81\pm0.36$ MeV~\cite{Cleo07:D-mass}. The
spectrum of the di-pion indicates that they come from the $\rho$
resonance and the charge parity of $X$ is even~\cite{Belle0309032}.
Moreover, analyzes both by Belle~\cite{Belle05:X3872:J^PC} and
CDF~\cite{CDF06:X3872:J^PC} favor the quantum number
$J^{PC}=1^{++}$.

The product branching ratio $\mathcal{B}(B^+\to
XK^+)\cdot\mathcal{B}(X\to J/\psi\rho(\pi^{+}\pi^{-}))$ is about
$7\mbox{-}10\times10^{-6}$~\cite{BaBar08:X3872:R_(0/+),Belle08:X3872:R_(0/+)}.
With the rate of this mode, the relative rates of other decay modes
of $X(3872)$
are\cite{Belle05:X3872:psi-omega,BaBar08:X3872:psi(')-gamma}
\ba R_{\psi\omega}&=&\frac{\mathcal{B}(X\to J/\psi~\omega)}{\mathcal{B}(X\to J/\psi~\rho)}=1.0\pm0.5,\label{R:psi-omega}\\
R_{\psi\gamma}&=&\frac{\mathcal{B}(X\to J/\psi~\gamma)}{\mathcal{B}(X\to J/\psi~\rho)}=0.33\pm0.12,\label{R:psi-gamma}\\
R_{\psi'\gamma}&=&\frac{\mathcal{B}(X\to
\psi'~\gamma)}{\mathcal{B}(X\to
J/\psi~\rho)}=1.1\pm0.4.\label{R:psi'-gamma} \ea

It is interesting that another narrow structure was found in the
$D^0\bar{D}^0\pi^0$~\cite{Belle06:X3872:DDpi} or the
$D^0\bar{D}^{*0}$~\cite{BaBar08:X3872:DD*} invariant mass spectrum
near 3875 MeV, which is a little higher than that in
(\ref{M:X3872}), in the decays $B^{+/0}\to
D^0\bar{D}^{*0}(\bar{D}^0\pi^0)K^{+/0}$. Recently,
Belle\cite{Belle08:X3872:DD*} updated the measurement on $X(3875)$
and improved their fitting method and found
\be M(X(3875))=3872.6^{+0.5}_{-0.4}\pm0.4~\mbox{MeV},\label{M:X3875}
\ee
which is consistent with that in (\ref{M:X3872}). Provided that the
two $X$-states are the same, one can extract the ratio
\be R_{DD^*}=\frac{\mathcal{B}(X\to D^0D^{*0})}{\mathcal{B}(X\to
J/\psi~\rho)}=9\pm2\label{R:DD*} \ee
from the Belle data~\cite{Belle08:X3872:DD*}.

The $X(3872)$ is widely accepted  as a molecule candidate of
$D^0\bar{D}^{*0}$ in
S-wave~\cite{Tornqvist:X3872:molecule,Swanson04:X3872:molecule}
since its mass is very close to the $D^0\bar{D}^{*0}$ threshold.
This assignment can also give a natural explanation of the $J^{PC}$
of $X(3872)$ and predict the ratio
$R_{\psi\omega}\sim1$~\cite{Swanson04:X3872:molecule}, which is in
agreement with that in (\ref{R:psi-omega}). However, as a loosely
bound state of $D^0\bar{D}^{*0}$, it should be difficult to be
produced in B-decays or in $p-p$ collision at the Tevatron. For
example, a model calculation~\cite{Braaten:X3872:B-production} shows
that in the $B^+$ decay a molecule $X(3872)$ has a branching ratio
$\mathcal{B}(B^+\to XK^+)=(0.07\mbox{-}1.0)\times10^{-4}$, whereas
the experimental rate tends to exceed this upper limit. Furthermore,
Belle Collaboration has observed $X(3872)$ in the neutral channel
$B^0\to X(J/\psi~\pi^+\pi^-)K^0$ with 5.9$\sigma$ significance and
with a rate almost as large as that of the charged
channel~\cite{Belle08:X3872:R_(0/+)}
\be \frac{\mathcal{B}(B^0\to X(3872)~K^0)}{\mathcal{B}(B^+\to
X(3872)~K^+)}=0.82\pm0.22\pm0.05, \ee
 which implies that $X(3872)$
is an isoscalar. The most serious problem of the molecular model, in
our opinion, is that it is difficult for a loosely bound state to
radiatively transit into exited charmonium, such as $\psi'$, through
quark annihilation or other mechanisms. Model
calculations~\cite{Swanson04:X3872:molecule} predict the ratio
\be R_{\psi'\gamma/\psi\gamma}=\frac{\mathcal{B}(X\to
\psi'~\gamma)}{\mathcal{B}(X\to \psi~\gamma)}\simeq
4\times10^{-3},\label{R:psi'gamma/psigamma:molecule} \ee
whereas the experimental value of this
ratio~\cite{BaBar08:X3872:psi(')-gamma} is
\be
R_{\psi'\gamma/\psi\gamma}^{ex}=3.4\pm1.4.\label{R:psi'gamma/psigamma:experimental}
\ee

Most of the above problems for the molecular model can be resolved
if one can assign $X(3872)$ as a conventional charmonium. As a
$J^{PC}=1^{++}$ state, the only candidate is the $\chi_{c1}'$ whose
mass is about 3.90 GeV in our model (see Table.I). The $30~MeV$
difference between the predicted mass and the experimental one in
(\ref{M:X3872}) can be further reduced if the coupled channel
effects are taken into account~\cite{Li08:Couppled-vs-Screened}. It
is the S-wave coupling of $\chi_{c1}'$ to $D^0\bar{D}^{*0}$ that
tends to lower the mass of $\chi_{c1}'$ towards the threshold of
$D^0\bar{D}^{*0}$. This is  related to the cusp effect at the
$D^0\bar{D}^{*0}$ threshold\cite{Bugg0802}.

The charmonium candidates of $X(3872)$ were
suggested~\cite{Barnes04:X3872:charmonium,Eichten04:X3872:charmonium}
soon after it was found. However, these suggestions were almost
given up, after the isospin-violating decay $X\to J/\psi\rho$ was
confirmed. Because of the coupled channel effects, the $\chi_{c1}'$
will mix with nearby opened $D^0\bar{D}^{*0}$ component. Such a
mixed charmonium model for $X(3872)$ was proposed in
Ref.\cite{Meng05:X3872:B-production} and
Ref.\cite{suzuki05:X3872:chi-c1-2P}. Differing from the molecular
models, the $D^0\bar{D}^{*0}$ component mixed in the $1^{++}$
charmonium is just a hadronic description for effects of the vacuum
polarization induced by the dynamical quark pair creation and
annihilation. Thus, the mixed chamonium is as compact as the
conventional charmonium. As a result, the production rates of
$X(3872)$ should be large and equal in both the neutral and charged
channels in $B$ meson decays~\cite{Meng05:X3872:B-production}. The
production rate of X(3872) in $p-p$ collisions at the Tevatron
should also be large, comparable to that of $\chi_{c1}(1P)$ (but
somewhat reduced due to a smaller $c\bar c$ norm in the mixed
charmonium model of X(3872)).

Using the final state rescattering mechanism, one may explain the
isospin violating decay $X(3872)\to
J/\psi~\rho$~\cite{Meng07:X3872:decays}. The isospin violation,
which is implied by the ratio $R_{\psi\omega}$ in
(\ref{R:psi-omega}),  is expected to be mainly due to the difference
between the thresholds of $D^0\bar{D}^{*0}$ and $D^+D^{*-}$, and the
larger phase space of $J/\psi~\rho$ than that of $J/\psi~\omega$
also favors the $J/\psi~\rho$
decay~\cite{suzuki05:X3872:chi-c1-2P,Meng07:X3872:decays}. In
addition, the ratio $R_{DD^*}$ in (\ref{R:DD*}) may also be
accounted for provided that the $X(3872)$ lies below or just a
little amount above the $D^0\bar{D}^{*0}$
threshold~\cite{Meng07:X3872:decays}.

The E1 transition rates of $\chi_{c1}'$ are sensitive to the
relativistic corrections due to the node in the $2P$ wave function,
especially for the one $\chi_{c1}'\to J/\psi~\gamma$. In our model,
after relativistic corrections are taken into account, the
transition widths $\Gamma(\chi_{c1}'\to
J/\psi(\psi')~\gamma)=45(60)$ KeV (see Table~\ref{E1rad}). The
corresponding ratio $R_{\psi'\gamma/\psi~\gamma}\simeq1.33$, which
is much larger than the one predicted by the molecular model in
(\ref{R:psi'gamma/psigamma:molecule}) and in rough agreement with
the experimental value (\ref{R:psi'gamma/psigamma:experimental}).
Different treatments or different parameters in the relativistic
corrections can result in very different estimations for the rate of
$\chi_{c1}'\to J/\psi~\gamma$ (2P-1S transition), while the rate of
$\chi_{c1}'\to \psi'~\gamma$ (2P-2S transition) can only be changed
a little. For example, Ref.~\cite{Barnes04:X3872:charmonium} gives
$\Gamma(\chi_{c1}'\to J/\psi(\psi')~\gamma)=11(64)$ KeV, and the
corresponding ratio $R_{\psi'\gamma/\psi~\gamma}\simeq6$. Thus, in
the mixed charmonium model for $X(3872)$, the expected range of the
ratio may be
\be
R_{\psi'\gamma/\psi\gamma}=1.3\mbox{-}6.0.\label{R:psi'gamma/psigamma:chi-c1'}
\ee

If we use the calculated $\Gamma(\chi_{c1}'\to \psi'\gamma)=60$ KeV
as input for $X(3872)\to\psi'\gamma$, and use the experimental
results  (\ref{R:psi-omega}), (\ref{R:psi-gamma}),
(\ref{R:psi'-gamma}), and (\ref{R:DD*}), as well as the width of
decay to light hadrons (assuming $\Gamma(\chi_{c1}(2P)\to ~light
~hadrons)\approx \Gamma(\chi_{c1}(1P)\to ~light ~hadrons)\approx
600~KeV$), we will get the total width of $X(3872)$ to be about
$1400\pm 300$~KeV, which is compatible with the measurement (it can
be further reduced when the $c\bar c$ norm in X(3872) is less than
one).

The nature of $X(3872)$ can also be uncovered by the pole structure
of the scattering amplitude involving the resonance near the
$D^0\bar D^{*0}$ threshold. This study is also needed to explain the
different peak locations of $X(3872)$ in the $J/\psi\pi^+\pi^-$ and
$D^0\bar D^0\pi^0/D^0\bar D^{*0}$ modes. Three
groups~\cite{hanhart07:X3872fit,braaten07:X3872fits,Zhang08:X3872fit}
have devoted themselves to this study and the conclusions are quite
different. One group~\cite{hanhart07:X3872fit} conclude that the
$X(3872)$ tends to be a virtual state of $D^0\bar D^{*0}$, while
another group's fit~\cite{braaten07:X3872fits} favors the loosely
bound state explanation. Most recently, with the Belle's new
data~\cite{Belle08:X3872:R_(0/+),Belle08:X3872:DD*}, authors of
Ref.\cite{Zhang08:X3872fit} gave a more systematic study on this
topic and found that there may need to be two near-threshold poles
to account for the data, one from the $D^0\bar D^{*0}$ component and
the other from the charmonium state $\chi_{c1}'$.

To sum up for the discussion in this subsection, we find that the
$\chi_{c1}(2P)$-dominated charmonium interpretation for the X(3872)
may account for (i) the E1 transition rates to $J/\psi$ and
$\psi(2S)$ and their ratio(\ref{R:psi'gamma/psigamma:experimental});
(ii) the large production rates in $B$ decays and equal rates for
$B^+$ and $B^0$; (iii) the large production rate in $p-p$ collisions
at the Tevatron; (iv) the isospin violating decay to $J/\psi\rho$.
Moreover, in the screened potential model the mass of
$\chi_{c1}(2P)$ is predicted to take a lower value than the quenched
potential model. However, though the mass of $\chi_{c1}(2P)$ can be
lowered by coupling to $D^0\bar D^{*0}$, one can not provide a
quantitative explanation for the extreme closeness of X(3872) to the
$D^0\bar D^{*0}$ threshold (say, within 0.5~MeV), which is the most
favorable motivation for the molecule interpretation.

\subsection{$X(3940), X(4160)$}

The $X(3940)$ was found by the Belle
Collaboration~\cite{Belle07:X3940} in the recoiling spectrum of
$J/\psi$ in the $e^+e^-$ annihilation process $e^+e^-\to J/\psi
+anything$ and $e^+e^-\to J/\psi +D\bar{D}^*$. The later was studied
further with higher statistics by Belle~\cite{Belle08:X3940&4160}.
The mass and width of $X(3940)$ are determined to be
\ba M(X(3940))&=&3942_{-6}^{+7}\pm6~\mbox{MeV},\label{m:X3940}\\
\Gamma(X(3940))&=&37_{-18}^{+26}\pm8~\mbox{MeV}.\label{Gamma:X3940}
\ea
Meanwhile, they also found the $X(4160)$ in the $D^*\bar{D}^*$ mode
in the process $e^+e^-\to J/\psi +D^*\bar{D}^*$ with a significance
of $5.1\sigma$. The mass and width of the $X(4160)$ are given by
\ba M(X(4160))&=&4156_{-20}^{+25}\pm15~\mbox{MeV},\label{m:X4160}\\
\Gamma(X(4160))&=&139_{-61}^{+111}\pm21~\mbox{MeV}.\label{Gamma:X4160}
\ea
Besides, there is a structure around 3880 MeV in the $D\bar{D}$
spectrum in $e^+e^-\to J/\psi +D\bar{D}$. However, it is too wide to
present a resonance shape sufficiently.

Both of the two $X$-states have large production rates in these
processes~\cite{Belle08:X3940&4160}. This fact implies that the
charge parities should be even since the charge odd state associated
with $J/\psi$ needs to be produced via two photon fragmentation,
which is expected to be highly suppressed~\cite{Chao08:X4160}. On
the other hand, the only known charmonium states that are produced
in this way are $\eta_c$, $\eta_c'$ and
$\chi_{c0}$~\cite{Belle04BaBar05:ee-psietac-psichic0},  and this
double charmonium production phenomenon can be explained in the
framework of nonrelativistic QCD \cite{doublecharmQCD}. The
production rates of $X(3940)$ and
$X(4160)$~\cite{Belle08:X3940&4160} are both as large as those of
$\eta_c$, $\eta_c'$ and $\chi_{c0}$. This suggests that the two
$X$-states could be either pseudoscalar like $\eta_c$ or scalar like
$\chi_{c0}$ (see Ref.~\cite{Chao08:X4160} for more detailed
discussions).

The observation that the dominant decay mode of $X(3940)$ being
$D\bar{D}^*$ and the lack of evidence for the $D\bar{D}$ decay mode
~\cite{Belle07:X3940,Belle08:X3940&4160} indicates that it can not
be a scalar but can be a good candidate for the $\eta_c(3S)$.  The
main problem is the low mass of $X(3940)$ as the $\eta_c(3S)$.
Although lower than that in the quenched potential
model~\cite{Barnes2005} by 50 MeV or more, the mass of $\eta_c(3S)$
in our screened potential model (see Table~\ref{spectrum_table}) is
still larger than the observed mass (\ref{m:X3940}) by about 50 MeV.
Moreover, the mass splitting between $X(3940)$ and $\psi(4040)$ is
larger than that between $\eta_c'$ and $\psi'$, which looks quite
unnatural. But it may be due to the coupled channel
effects~\cite{Eichten06:X3940}, which will further lower the
$\eta_c(3S)$ mass hopefully.

The dominant decay mode of $X(4160)$ is
$D^*\bar{D}^*$~\cite{Belle08:X3940&4160}, and the other modes, such
as $D\bar{D}$ and $D\bar{D}^*$, were not seen. Thus the charmonium
candidates can be $\eta_c(4S)$ or $\chi_{c0}(3P)$, whose masses are
4250 MeV and 4131 MeV in our model prediction, respectively.
Evidently, the mass of $X(4160)$ in (\ref{m:X4160}) favors the
$\chi_{c0}(3P)$ candidate. The $\chi_{c0}(3P)$ can not decay into
$D\bar{D}^*$, and the decay $\chi_{c0}(3P)\to D\bar{D}$ is expected
to be strongly suppressed by the form factor and the effects induced
by the nodes of the $3P$ wave function, just like the case of
suppressed $\psi(4040)\to D\bar{D}$ decay~\cite{Barnes2005}. The
main problem of this assignment may be why the $\chi_{c0}(2P)$ state
is not found in the similar process. One possible
account~\cite{Chao08:X4160} is that the broad peak around 3880 MeV
in the $D\bar{D}$ spectrum~\cite{Belle08:X3940&4160} mentioned above
could be the missing $\chi_{c0}(2P)$ state, since its mass in our
model is about 3842 MeV and just lies within the bump (note,
however, that this bump might not be a resonance\cite{Bugg0811}). In
addition, the measurements on angular distributions can be used to
test the two possible assignments, $\eta_c(4S)$ and $\chi_{c0}(3P)$,
for the $X(4160)$~\cite{Chao08:X4160}.

\section{Summary and Conclusions}

In this paper, we try to incorporate the color-screening (string
breaking) effect due to light quark pair creation into the heavy
quark-antiquark long-range confinement potential, and investigate
the effects of screened potential on the spectrum of the charmonium
especially the higher charmonium. We calculate the masses,
electromagnetic decays, and E1 transitions of charmonium states in
the nonrelativistic screened potential model, and propose possible
assignments for the newly discovered charmonium or charmonium-like
states, i.e., the so-called $"X,Y,Z"$ mesons. We find that compared
with the unscreened potential model, the masses predicted in the
screened potential model are considerably lower for higher
charmonium states. For example, the predicted $\chi_{c2}(2P)$ mass
well agrees with that of the Z(3930), and the mass of $\psi(5S)$
rather than $\psi(4S)$ is compatible with that of $\psi(4415)$. As a
result of the compressed mass spectrum in our model,  most of the
$X,Y,Z$ states might be accomodated in the conventional higher
charmonia. In particular, the discovered four $Y$ states in the ISR
process, i.e., $Y(4008), Y(4260), Y(4320/4360), Y(4660)$ may be
assigned as the $\psi(3S), \psi(4S), \psi(3D), \psi(6S)$ states
respectively. The $X(3940)$ and $X(4160)$ found in the double
charmonium production in $e^+e^-$ annihilation may be assigned as
the $\eta_c(3S)$ and $\chi_{c0}(3P)$ states respectively. Based on
the calculation of E1 transition widths for $\chi_{c1}(2P)\to \gamma
J/\psi$ and $\chi_{c1}(2P)\to \gamma \psi(2S)$ and other results, we
argue that the $X(3872)$ may be a $\chi_{c1}(2P)$ dominated
charmonium state with some admixture of the $D^0\bar{D}^{*0}$
component. The problems encountered in these assignments and
comparisons with other interpretations for these $X,Y,Z$ mesons are
discussed in detail. We emphasize that more theoretical and
experimental investigations are urgently needed to clarify these
assignments and other interesting interpretations. In particular, we
hope experiments at BESIII and SuperBelle in the future will be
crucially useful in searching for new hadrons including
charmonium-like states and testing the theoretical interpretations.

\section{Acknowledgement}
We would like to thank Ce Meng for many valuable discussions and
assistance during this work, and Han-Qing Zheng and Shi-Lin Zhu for
useful discussions. We also thank D. Bugg for helpful comments. This
work was supported in part by the National Natural Science
Foundation of China (No 10675003, No 10721063).

\begin{table*}
\caption{Experimental and theoretical mass spectrum of charmonium
states. The experimental masses are PDG~\cite{PDG08} averages. The
masses are in units of MeV, while the averaged radii are in units of
$\rm fm$. The results of our screened potential model are shown in
comparison with that of Ref.\cite{Barnes2005} including the NR and
GI models\cite{Barnes2005}.}
 \vspace{0.5cm}
\begin{tabular}{cc|c|cc|cc}
\hline
 \multicolumn{2}{c|}{State} &Expt.&\multicolumn{2}{c|}{Theor. of ours}& \multicolumn{2}{c}{Theor. of Ref.\cite{Barnes2005}}\\
   &                          &         & \hspace{0.3cm}Mass\hspace{0.3cm}  & \hspace{0.3cm}$\langle r^2\rangle^{\frac{1}{2}}$\hspace{0.3cm} & \hspace{0.3cm}NR \hspace{0.3cm} &   GI \\
\hline
1S &  $J/\psi(1^3{\rm S}_1) $ &  $ 3096.916\pm 0.011$ & 3097           &0.41                   & 3090 & 3098 \\
   &  $\eta_c(1^1{\rm S}_0) $ &  $ 2980.3 \pm 1.2   $ & 2979           &                       & 2982 & 2975 \\
\hline
2S &  $\psi'(2^3{\rm S}_1)  $ &  $ 3686.093\pm 0.034$ & 3673           &0.91                   & 3672 & 3676 \\
   &  $\eta_c'(2^1{\rm S}_0)$ &  $ 3637 \pm 4       $ & 3623           &                       & 3630 & 3623 \\
\hline
3S &  $\psi(3^3{\rm S}_1)   $ &  $ 4039 \pm 1       $ & 4022           &1.38                   & 4072 & 4100 \\
   &  $\eta_c(3^1{\rm S}_0) $ &                       & 3991           &                       & 4043 & 4064 \\
\hline
4S &  $\psi(4^3{\rm S}_1)   $ &  $ 4263^{+8}_{-9}   $ & 4273           &1.87                   & 4406 & 4450 \\
   &  $\eta_c(4^1{\rm S}_0) $ &                       & 4250           &                       & 4384 & 4425 \\
\hline
5S &  $\psi(5^3{\rm S}_1)   $ &  $ 4421 \pm 4       $ & 4463           &2.39                   &      &      \\
   &  $\eta_c(5^1{\rm S}_0) $ &                       & 4446           &                       &      &      \\
\hline
6S &  $\psi(6^3{\rm S}_1)   $ &  $                  $ & 4608           &2.98                   &      &      \\
   &  $\eta_c(6^1{\rm S}_0) $ &                       & 4595           &                       &      &      \\
\hline
1P &  $\chi_2(1^3{\rm P}_2) $ &  $ 3556.20 \pm 0.09 $ & 3554           &0.71                   & 3556 & 3550 \\
   &  $\chi_1(1^3{\rm P}_1 )$ &  $ 3510.66 \pm 0.07 $ & 3510           &                       & 3505 & 3510 \\
   &  $\chi_0(1^3{\rm P}_0) $ &  $ 3414.75 \pm 0.31 $ & 3433           &                       & 3424 & 3445 \\
   &  $h_c(1^1{\rm P}_1)    $ &  $ 3525.93 \pm 0.27 $ & 3519            &                       & 3516 & 3517 \\
\hline
2P &  $\chi_2(2^3{\rm P}_2) $ &  $3929\pm5\pm2      $ & 3937           &1.19                   & 3972 & 3979 \\
   &  $\chi_1(2^3{\rm P}_1) $ &                       & 3901           &                       & 3925 & 3953 \\
   &  $\chi_0(2^3{\rm P}_0) $ &                       & 3842           &                       & 3852 & 3916 \\
   &  $h_c(2^1{\rm P}_1) $    &                       & 3908           &                       & 3934 & 3956 \\
\hline
3P &  $\chi_2(3^3{\rm P}_2) $ &                       & 4208           &1.67                   & 4317 & 4337 \\
   &  $\chi_1(3^3{\rm P}_1) $ &                       & 4178           &                       & 4271 & 4317 \\
   &  $\chi_0(3^3{\rm P}_0) $ &                       & 4131           &                       & 4202 & 4292 \\
   &  $h_c(3^1{\rm P}_1) $    &                       & 4184           &                       & 4279 & 4318 \\
\hline
1D &  $\psi_3(1^3{\rm D}_3) $ &                       & 3799           &0.96                   & 3806 & 3849 \\
   &  $\psi_2(1^3{\rm D}_2) $ &                       & 3798           &                       & 3800 & 3838 \\
   &  $\psi(1^3{\rm D}_1) $   &  $ 3775.2 \pm 1.7 $   & 3787           &                       & 3785 & 3819 \\
   &$\eta_{c2}(1^1{\rm D}_2)$ &                       & 3796           &                       & 3799 & 3837 \\
\hline
2D &  $\psi_3(2^3{\rm D}_3) $ &                       & 4103           &1.44                   & 4167 & 4217 \\
   &  $\psi_2(2^3{\rm D}_2) $ &                       & 4100           &                       & 4158 & 4208 \\
   &  $\psi(2^3{\rm D}_1) $   &  $  4153\pm3      $   & 4089           &                       & 4142 & 4194 \\
   &$\eta_{c2}(2^1{\rm D}_2)$ &                       & 4099           &                       & 4158 & 4208 \\
\hline
3D &  $\psi_3(3^3{\rm D}_3) $ &                       & 4331           &1.94                    &      &      \\
   &  $\psi_2(3^3{\rm D}_2) $ &                       & 4327           &                       &      &      \\
   &  $\psi(3^3{\rm D}_1) $   &  $                $   & 4317           &                       &      &      \\
   &$\eta_{c2}(3^1{\rm D}_2)$ &                       & 4326           &                       &      &      \\

\hline \hline
\end{tabular}
\label{spectrum_table}
\end{table*}

\begin{table*}
\caption{Leptonic widths (in units of KeV) for charmonium states
without S-D mixing in the screened potential model. The widths
calculated with and without QCD corrections are marked by
$\Gamma_{ee}$ and $\Gamma^{0}_{ee}$ respectively. The experimental
values are taken from PDG~\cite{PDG08}.}
 \begin{center}
 \begin{tabular}{|@{\hspace{0.5cm}}c@{\hspace{0.5cm}}|@{\hspace{0.5cm}}c@{\hspace{0.5cm}}|@{\hspace{0.5cm}}c@{\hspace{0.5cm}}|@{\hspace{0.5cm}}l@{\hspace{0.5cm}}|}
 \hline
 state             &$\Gamma^{0}_{ee}$ &$\Gamma_{ee}$ &$\Gamma^{expt}_{ee}$\\
 \hline
 $1{}^3S_{1}(3097)$       &11.8              &6.60 &$5.55\pm0.14\pm0.02$\\
 \hline
 $2{}^3S_{1}(3686)$       &4.29              &2.40 &$2.33\pm0.07$\\
 \hline
 $3{}^3S_{1}(4039)$       &2.53              &1.42 &$0.86\pm0.07$\\
 \hline
 $4{}^3S_{1}(4263)$ &1.73              &0.97& \\
 \hline
 $5{}^3S_{1}(4421)$       &1.25              &0.70 & $0.58\pm0.07$\\
\hline
 $6{}^3S_{1}(4664)$       &0.88              &0.49 & \\
 \hline
 $1{}^3D_{1}(3775)$       &0.055             &0.031 &$0.259\pm0.016$\\
 \hline
 $2{}^3D_{1}(4153)$       &0.066             &0.037 &$0.83\pm0.07$\\
\hline
 $3{}^3D_{1}(4361)$       &0.079             &0.044 &\\
 \hline
\end{tabular}
\end{center}
\label{tab:ee}
\end{table*}

\begin{table*}
\caption{Two-photon decay widths (in units of KeV) of pseudoscalar
  (${}^1\!S_0$), scalar (${}^3\!P_0$), and tensor (${}^3\!P_2$) charmonium states. Charmonium masses are in units of
  MeV.}
\begin{ruledtabular}
\begin{tabular}{c|c|ccccccc |c  }
      &                    & \multicolumn{7}{c|}{Theory}                                                                                  & Experiment\\
\hline
 state&  mass              &Ref.\cite{Ebert0302044} &Ref.\cite{Munz:1996hb} &Ref.\cite{gjr}&Ref.\cite{sbg}&Ref.\cite{hlt} &Ref.\cite{ab} &Ours   &PDG~\cite{PDG08} \\
\hline
 $\eta_c(1{}^1\!S_0)$&2980 & 5.5                    & 3.5                   &10.94         & 7.8          & 5.5        &4.8        & 8.5  &$6.7^{+0.9}_{-0.8}$\\
\hline
 $\eta'_c(2{}^1\!S_0)$&3637&1.8                     & 1.38                  &              & 3.5          & 2.1        & 3.7       &2.4   & \\
\hline
$\eta'_c(3{}^1\!S_0)$&3991 &                        & 0.94                  &              &              &            &           &0.88  & \\
\hline
 $\chi_{c0}(1{}^3\!P_0)$&3415&2.9                   &1.39                   &6.38          &2.5           &5.32        &           &2.5  &$2.40\pm0.29$ \\
\hline
$\chi'_{c0}(2{}^3\!P_0)$&3842&1.9                   &1.11                   &              &              &            &           &1.7  & \\
\hline
$\chi'_{c0}(3{}^3\!P_0)$&4156&                      &0.91                   &              &              &            &           &1.2  & \\
\hline
 $\chi_{c2}(1{}^3\!P_2)$&3556&0.50                  &0.44                   &0.57          &0.28          &0.44        &           &0.31  &$0.49\pm0.05$\\
\hline
$\chi'_{c2}(2{}^3\!P_2)$&3929&0.52                  &0.48                   &              &              &            &           &0.23  & \\
\hline
$\chi'_{c2}(3{}^3\!P_2)$&4208&                      &0.014                  &              &              &            &           &0.17  & \\
\end{tabular}
\end{ruledtabular}
\label{twophoton}
\end{table*}

\begin{table*}
\caption{E1 transition rates of charmonium states in the
non-screened potential model~\cite{Barnes2005}~(marked by $NR$) and
our screened potential model~(those calculated by the zeroth-order
wave functions are marked by $SNR_0$ and those by the first-order
relativistically corrected wave functions are marked by $SNR_1$).}
\vskip 0.3cm
\begin{ruledtabular}
\begin{tabular}{l| ll| c c| c c c| c c  }
state & Initial meson  & Final meson
&\multicolumn{2}{c}{E$_{\gamma}$ (MeV)} &
\multicolumn{3}{c}{$\Gamma_{\rm thy}$~(keV)}
& $\Gamma_{\rm expt}$~(keV) & \\
&                &             & NR  & $SNR_{0(1)}$  & NR  & $SNR_0$&$SNR_1$  \\
\hline
2S $\to$ 1P    &$\psi'(2^3{\rm S}_1)(3686)$& $\chi_{c2}(1^3{\rm P}_2)$   &128    &128     &38   &43  &34 & 26.3 $\pm$ 1.5   \\

               &                           & $\chi_{c1}(1^3{\rm P}_1)$  & 171    & 171    &54   &62  &36 & 27.9 $\pm$ 1.5   \\

               &                           & $\chi_{c0}(1^3{\rm P}_0)$  & 261    & 261    &63   &74  &25 & 29.8 $\pm$ 1.5   \\
               &$\eta_c(2^1{\rm S}_0)(3637)$&$h_c(2^1{\rm P}_1)$        &        & 109    &     &146 &104&\\
\hline
1P $\to$ 1S    &$\chi_{c2}(1^3{\rm P}_2)(3556)$&$J/\psi(1^3{\rm S}_1)(3097)$  & 429    &429     & 424 &473 &309& 406 $\pm$ 31   \\
               & $\chi_{c1}(1^3{\rm P}_1)(3511)$&                             & 390    &390     & 314 &354 &244& 320 $\pm$ 25   \\
               & $\chi_{c0}(1^3{\rm P}_0)(3415)$&                             & 303    &303     & 152 &167 &117& 131 $\pm$ 14   \\
               & $h_c(1^1{\rm P}_1)(3525)$      & $\eta_c(1^1{\rm S}_0)(2980)$& 504    &504     & 498 &764 &323&  \\
\hline
2P $\to$ 1S    &$\chi_{c2}(2^3{\rm P}_2)(3929) $ &$J/\psi(1^3{\rm S}_1) $  & 779    &744     & 81 &101 &109&    \\
               & $\chi_{c1}(2^3{\rm P}_1)(3872) $&                            & 741    &697     & 71 &83  &45&   \\
               & $\chi_{c0}(2^3{\rm P}_0)(3842)$ &                                & 681    &672     & 56 &74  &9.3&    \\
               & $h_c(2^1{\rm P}_1)(3908)$       & $\eta_c(1^1{\rm S}_0)$   & 839    &818     & 140&134 &250&  \\

\hline
2P $\to$ 2S    &$\chi_{c2}(2^3{\rm P}_2) $ &$J/\psi(2^3{\rm S}_1)$& 276    &235     & 304 &225 &100&    \\
               & $\chi_{c1}(2^3{\rm P}_1) $&                            & 232    &182     & 183 &103 &60&    \\
               & $\chi_{c0}(2^3{\rm P}_0)$ &                                & 162    &152     & 64  &61  &44&    \\
               & $h_c(2^1{\rm P}_1)$       & $\eta_c(2^1{\rm S}_0)$     & 285    &261     & 280 &309 &108&  \\

\hline
 1D $\to$ 1P   & $\psi_3(1^3{\rm D}_3)(3799) $   & $\chi_{2}(1^3{\rm P}_2)$   & 242    &236     & 272 &284 &223&\\
               & $\psi_2(1^3{\rm D}_2)(3798) $   & $\chi_{c2}(1^3{\rm P}_2) $ & 236    &234     &  64 &70  &55 &\\
               &                           & $\chi_{c1}(1^3{\rm P}_1)$  & 278    &276     & 307 &342 &208&\\
               & $\psi(1^3{\rm D}_1)(3775) $     & $\chi_{c2}(1^3{\rm P}_2)$  & 208    &213     & 4.9 &5.8 &4.6& $ <21$~\cite{Briere:2006ff} \\
               &                           & $\chi_{c1}(1^3{\rm P}_1)$  & 250    &255     & 125 &150 &93&$70\pm17$~\cite{Briere:2006ff} \\
               &                           & $\chi_{c0}(1^3{\rm P}_0)$  & 338    &343     & 403 &486 &197& $172\pm30$~\cite{Briere:2006ff} \\
               & $h_{c2}(1^1{\rm D}_2)(3796) $   & $h_c(1^1{\rm P}_1)$        & 264    &260     & 339 &575 &375&\\
\hline
\end{tabular}
\end{ruledtabular}
\label{E1rad}
\end{table*}

\end{document}